\documentclass[
]{ceurart}

\usepackage{listings}
\begin{document}

\copyrightyear{2022}
\copyrightclause{Copyright for this paper by its authors.
  Use permitted under Creative Commons License Attribution 4.0
  International (CC BY 4.0).}

\conference{PASIR'22: First Workshop on Proactive and Agent-Supported Information Retrieval at CIKM 2022,
  October 21, 2022,  Atlanta, GA}

\title{Towards Proactive Information Retrieval in Noisy Text with Wikipedia Concepts}

% \tnotetext[1]{You can use this document as the template for preparing your
%   publication. We recommend using the latest version of the ceurart style.}

\author{Tabish Ahmed}[%
email=tabish.ahmed.21@ucl.ac.uk,
]
% \cormark[1]
\address[]{Centre for Artificial Intelligence, University College London, UK.}

\author{Sahan Bulathwela}
% [%
% orcid=0000-0001-7116-9338,
% email=i.tiddi@vu.nl,
% url=https://kmitd.github.io/ilaria/,
% \address[3]{Vrije Universiteit Amsterdam, De Boelelaan 1105, 1081 HV Amsterdam, The Netherlands}

% \author[4]{Manfred Jeusfeld}[%
% orcid=0000-0002-9421-8566,
% email=Manfred.Jeusfeld@acm.org,
% url=http://conceptbase.sourceforge.net/mjf/,
% \address[4]{University of Skövde, Högskolevägen 1, 541 28 Skövde, Sweden}

% %% Footnotes
% \cortext[1]{Corresponding author.}
% \fntext[1]{These authors contributed equally.}

\begin{abstract}
Extracting useful information from the user history to clearly understand informational needs is a crucial feature of a proactive information retrieval system. Regarding understanding information and relevance, Wikipedia can provide the background knowledge that an intelligent system needs. This work explores how exploiting the context of a query using Wikipedia concepts can improve proactive information retrieval on noisy text. We formulate two models that use entity linking to associate Wikipedia topics with the relevance model. Our experiments around a podcast segment retrieval task demonstrate that there is a clear signal of relevance in Wikipedia concepts while a ranking model can improve precision by incorporating them. We also find Wikifying the background context of a query can help disambiguate the meaning of the query, further helping proactive information retrieval.
\end{abstract}

% \begin{keywords}
%   LaTeX class \sep
%   paper template \sep
%   paper formatting \sep
%   CEUR-WS
% \end{keywords}

\maketitle

\section{Introduction}

The informational needs of people are highly contextual and can depend on many different factors such as their current knowledge state, interests and goals \cite{trueeducation,goal_based_edrec,bulathwela2022power}. However, an effective information retrieval companion should minimise the human effort required in i) expressing a human information need and ii) navigating a lengthy result set. \textcolor{black}{Using topical representations of the user history (e.g. \cite{truelearn}) can immensely help formulating zero shot queries and refining short user queries that enable proactive information retrieval (IR).} 
% With the information overload, both of the above tasks have become challenging posing struggles in creating AI systems that can cooperate with human users to support their informational journey. 
% With the emergence of the semantic web, the majority of digital information in today’s world is enriched with metadata that is easily manageable with machines. 
% Developing Artificial Intelligence (AI) that can mildly, at least understand how information and knowledge is structured can help address this struggle. 
{While the world has digital textual information in abundance, it can often be noisy (e.g. extracted through Automatic Speech Recognition (ASR), PDF text extraction etc.), leading to state-of-the-art neural models being highly sensitive to the noise producing sub-optimal results \cite{sidiropoulos2022impact}. This demands denoising steps to refine both query and document representation.}

{In this paper, we argue that Wikipedia, an openly available encyclopedia, can be a humanly intuitive knowledge base \cite{democratise2021} that has the potential to provide the world view many \emph{noisy} information Retrieval systems need. In the midst of noisy text, we point out that it can help refine user queries (both explicit and implicit) while compiling large result sets into structured narratives that human users can navigate with optimal effort. We use a dataset with noisy text where the query description is used as a proxy of the user's context used in a proactive information retrieval system. We demonstrate that positive results can be obtained in comparison to relevant baselines when ranking documents.}

\section{Related Work} \label{sec:lit}

Retrieving relevant documents to a personal information need involves mining textual features that can be used to match them to user queries.
% \subsection{Retrieving Relevant Documents}
% Many different approaches in information retrieval rely on textual features of the user queries and the documents to compute relevance. 
\textcolor{black}{While previous works have explored Wikipedia-based concept modelling for query expansion \cite{azad2019new, nasir2019knowledge} exclusively, our work deviates from these works as we focus on i) simultaneously adding Wiki features to the documents, ii) ranking documents with noisy texts and iii) using the query description (proxy for context extracted from user history) for query disambiguation.} 
\textcolor{black}{Latest neural ranking models dominate the state-of-the-art in IR albeit neural models struggle in the presence of noise in text \cite{sidiropoulos2022impact} making much simpler, computationally efficient IR models such as BM25, DPH and PL2 much more suitable for noisy data \cite{jones2021trec}. 
Prior evidence has demonstrated the success of simpler models such as BM25 with the exception of failure on long documents \cite{lv2011documents}. However, more sophisticated probabilistic models such as DPH \cite{amati2006frequentist}, motivated by the Divergence from Randomness (DFR) framework based on a hypergeometric distribution and  Popper normalisation instead of Laplace normalization, have outperformed the BM25 model when it comes to noisy text \cite{jones2021trec}.}
% Our motivation to use simple architectures such as BM25, DPH, PL2 as juxtaposed the recent state-of-the-art architectures as baselines in \cite{azad2019new} as baselines can be found in the results published in \cite{jones2021trec} and a more recent corroboration of the same by \cite{sidiropoulos2022impact}.
% \textcolor{red}{Furthermore, a key deviation of our approach is not only to enrich the queries, but also to enrich the noisy segments and investigate if it can be a suitable means to increase retrieval effectiveness.} 

% Probabilistic relevance models that rely on the bag-of-words representation of documents and queries such as BM25 \cite{robertson2009probabilistic} are foundational to information retrieval.

% The DPH model in Python-Terrier is an adaption of 
% The models differ in approach, BM25 relies on probabilistic retrieval in a parametric way whereas DPH relies on frequentist approach in a non-parametric way. A linear combination of the two scores provides gains in performance after the Query is expanded using entities from the query descriptions.  
% The same model is used without entities but expansion using Wikipedia Concepts for a direct comparison with the NER approach. Later the model a new model is proposed where we expand the Queries using the Wikipedia concepts from Queries+Descriptions as well as entities from descriptions. 

This emerging need for retrieving from noisy-text has led to recent competitions such as the podcast segment retrieval challenge \cite{yu2020spotify} that has paved way to improving the state-of-the-art. \textcolor{black}{This competition is relevant to our work as we also tackle proactive IR in noisy text.}
% While more sophisticated machine learning based  and deep learning-based ranking models \cite{rank_svm,RankNET_fam} have emerged over time, it is still evident that simple count-based ranking algorithms still dominate solution tables \cite{jones2021trec} \textcolor{red}{in context of noisy context where employing end-to-end learning can rather see a drop in performance \cite{sidiropoulos2022impact}}. 
% This is due to multiple reasons, The probabilistic bag-of-words methods are word token based, making them significantly more explainable and interpretable in comparison to their complex counterparts. The computation cost of the simpler probabilistic models is also substantially lower, usually only having to conduct hyperparameter tuning and avoid training altogether. Such computationally efficient models are also much simpler to operate in real world scenarios attracting the majority of the community to still use them. 
While the best performing model within the competition uses a neural approach, it uses token embeddings rather than document embeddings \cite{galuvscakova2020combine}. The leading model uses a re-ranking model trained on an orthogonal dataset adding to computational costs significantly. The next best, Dublin City University model (DCU) uses a variety of features ranging from text, WordNet based synonyms and entities for expanding the query \cite{moriya2020dcu}. As the DCU model uses automatically extracted entities (via Named Entity Recognition) to expand the query, we consider this work as the most relevant model to our proposal. But, we also annotate the segments with Wikipedia concepts/entities further differentiating our work from theirs.
User queries, inherently are short, averaging two to three key words in general \cite{arampatzis2008study}. Exploiting auxiliary information available to recover the context of the search is essential in such scenarios. In recent competitions such as the podcast segment retrieval challenge, the queries used were accompanied by a small description that provided the contextual information about the query itself. It was evidenced that the leading models in this challenge had to use the description to extract features and sometimes use the identified keywords to make queries to external search engines to further expand the queries \cite{jones2021trec}. However, in a real world system, such information is not available. Hence a proactive IR system has to rely on the user interactions from the past to build the context \cite{sen2021proactive}. In this work, we use the query context as a proxy representation extracted by a proactive IR system.

\emph{Proactive information retrieval} is a field in IR that gained a lot of attention recently. Supporting user information queries with minimal/zero effort will enable users to use IR systems more effectively. However, a key part of facilitating information retrieval tasks proactively is building systems that can capture the relevant signal from the user that indicate the informational needs of them. Prior works in this domain have identified several ways to harvest user intent. The simplest approach in this direction is to explicitly ask questions \cite{torbati2020personalized} or use the demographic information of the user \cite{yang2016modeling} to profile them. Rather than explicitly disrupting the user experience, many works harvest implicit user actions such as clicks, impressions and previous searches \cite{peek_orsum,yang2016modeling}. Especially in the social media domain, the interactions of users are used to create concept-based models \cite{zarrinkalam2020extracting} that can be used to infer user preferences and create zero effort queries for new content. The extracted preference features are usually used to expand information queries \cite{nasir2019knowledge,azad2019new}, synthesise zero-effort queries \cite{zarrinkalam2020extracting} or personalise search results \cite{Syed2017}. In this work, we argue in support of using concept-based user models as a form of zero-effort query for informational recommendation and learning scenarios, especially in the presence of noisy document representations.

\subsection{Concept-based User Modelling and Wikipedia}

% While concept-based user modelling can be utilised to compose zero-effort queries using user past interactions, there are many definitions for a \emph{concept} in this application domain. 
In the field of document retrieval and personalisation, keyword extraction is a heavily used form of identifying \emph{concepts} from textual documents. When working with short text such as social media posts, the common method is to use the words in the posts as words and build a user profile over these words \cite{piao_mooc}. While this approach provides fine grained features, the feature space can be too vast and pose challenges in increasing recall. To address this issue, other systems use topic detection using unsupervised learning approaches such as LDA \cite{zarrinkalam2017predicting}. Such unsupervised techniques are complex to tune and are not guaranteed to give humanly intuitive topics. While expert annotation (e.g. in education \cite{Corbett1994}) has been one of the approaches to obtain humanly intuitive, representative topics from textual content, this approach is not scalable. \textbf{{Wikification}}, a more recent approach, looks promising towards automatically extracting explainable concepts from text. Wikification identifies Wikipedia concepts present in a document by connecting natural text to Wikipedia articles via entity linking \cite{wikifier}. In this work, we argue for the usefulness of Wikification in creating precise, humanly intuitive concept representations at scale. Our experiments also demonstrate how Wikipedia concepts can be used to i) disambiguate query keyword and ii) at the absence of them, create entity features that can find relevant documents for users proactively.

\section{Wikipedia to Support Proactive Search-based Systems} \label{sec:task}

Through this work, we aim to explore how the knowledge contained in Wikipedia can be utilised to better a proactive search-based system. In order to tread towards this direction, we aim to answer multiple research questions in this work as a promising step. We hypothesise that associating Wikipedia concepts to a natural language based IR system can improve a proactive IR system by i) bettering disambiguation of the meaning of user queries, ii) by exploiting the semantic meaning of document words while being less sensitive to noisy words in the document transcripts and iii) by creating opportunities for the system to understand the learner interactions more meaningfully.

\begin{description}
    \item[RQ1] Do Wikipedia concepts carry a signal that indicates relevance of documents to queries? 
    \item[RQ2] Can Wikipedia annotations improve noisy information retrieval?
\end{description}

Through this work, we conduct a few preliminary experiments to validate these research questions.
% \textcolor{red}{Here you need to explain what are the experiments ran. You say that the segment retreival task in the podcaset TREC competition was use to bench mark the effectiveness of Wikipedia topics. tell that you are using pyterrier setup to execute the search experiments.}
We choose the segment retrieval task from the Podcast Track at TREC 2020 \cite{yu2020spotify} for these experiments as ''search'' is a key component of proactive information retrieval. We use this task specifically as the query description is provided as part of the query, providing the context of the query. In many proactive IR systems, the query context is mined using the user interaction history. In this case, the description can be treated as a context representation mined from the users' prior interactions.    

% first by annotating both the documents and queries with Wikipedia concepts using the Wikifier \cite{brank2017annotating} API and then 

% using Python Terrier to use different baselines i) BM25 and  ii) DPH \cite{amati2006frequentist} models to check the retrieval performance and then proceeding to replicate a state-of-the-art approach from TREC 2020 submissions by Dublin City University\cite{moriya2020dcu} which QE using the provided query descriptions using NER from spaCy. We append the Wikipedia Concepts in three incremental steps for the documents i) by using the top 1,3,5,10,20 concepts according to the Cosine Similarity Score ii) by using the top 1,3,5,10,20 concepts according to the PageRank score and iii) by using all concepts in the documents without any selection criterion. For the queries, we followed a rather simpler approach where used the query text as well as the description to annotate them with the Wikipedia concepts.

\subsection{Information Retrieval Models} \label{sec:models} 
% \textcolor{red}{In this section, you describe the models that you used in the experiments. You should reprt BM25, DPH and Yasufumi models as baselines. We then propose Wikified + entity and Tabish model as new models.You need to describe these models, how they work.}

We restrict our model choices to a probabilistic retrieval method BM25 and a hypergeometric weighing model DPH. The reasons for restricting to probabilistic models are i) its superior performance in noisy IR \cite{jones2021trec}, ii) the simplicity to introducing Wikipedia concept features, iii) the data and computational efficiency in training models and iv) the practical feasibility to real world systems. The models we utilise in this these experiments use the query $q$, query description $d$ and the segment $s$. 

\subsubsection{Baseline Models} 
We use BM25 and DPH models as probabilistic baseline models for this work. These two models use the textual tokens of the query $q$ and the segment $s$ to compute the relevance score $rel$ as per equation \ref{eq:bm25} where $\texttt{txt}$ identifies the text token features..

\begin{equation} \label{eq:bm25}
\centering
    rel(q,s) = f(q_\texttt{txt}, s_\texttt{txt})
\end{equation}

As described in section \ref{sec:lit}, we implement and test the DCU run 2 model \cite{moriya2020dcu} (referred to as DCU model hereafter) that expands the query with entities extracted from the description as per equation \ref{eq:dcu} that expands equation \ref{eq:bm25}. The entities are extracted using Named Entity Recognition (NER). 

\begin{equation} \label{eq:dcu}
\centering
    rel(q,\textcolor{blue}{d},s) = f(q_\texttt{txt} \textcolor{blue}{+ d_\texttt{ent}}, s_\texttt{txt})
\end{equation}

where \texttt{ent} represents the entities extracted from the query description $d$.

\subsection{Proposed Models}

The proposed models in this work uses Wikipedia concepts to enrich the representations used in the IR model. We hypothesise that adding Wikipedia concepts can improve precision of the model as exact entities can be matched between the query + context and the segments. We propose two models that use Wikipedia topics both in query and segment expansion. 

\paragraph{Wiki\_rel}, This model expands the query and the document by adding the Wikipedia concepts extracted using both the query $q$ and the query description $d$ as per equation \ref{eq:wiki}. This model replaces the entities in the DCU model by Wikipedia concepts.

\begin{equation} \label{eq:wiki}
\centering
    rel(q,d,s) = f(q_\texttt{txt}\textcolor{orange}{+\ q_\texttt{wiki}} \textcolor{orange}{+\ d_\texttt{wiki}}, s_\texttt{txt} \textcolor{orange}{+\ s_\texttt{wiki}})
\end{equation}

where $\cdot_{\texttt{wiki}}$ represents the concepts extracted from the query $q$, description $d$ and segment $s$ .

\paragraph{Ent\_Wiki\_rel}, This model expands the DCU model by enriching the query, description and the segment with Wikipedia concepts. This formulation is presented in equation \ref{eq:wiki}. 

\begin{equation} \label{eq:wiki}
\centering
    rel(q,d,s) = f(q_\texttt{txt} \textcolor{blue}{+ d_\texttt{ent}} \textcolor{orange}{+\ q_\texttt{wiki}} \textcolor{orange}{+\ d_\texttt{wiki}}, s_\texttt{txt} \textcolor{orange}{+\ s_\texttt{wiki}})
\end{equation}

The two proposed models help us validate RQ2 by i) Wiki\_rel model validating if \emph{replacing} named-entities with Wikipedia features improves the ranking models and ii) Wiki\_Ent\_rel model validating if \emph{adding} Wikipedia features improves the model.  

% To validate RQ1 and RQ2 as juxtaposed to the NER approach with QE using spaCy we use a linear combination of the scores from BM25 and DPH similar to what was used by DCU. We then try a simple retrieval exercise using Wikified Queries and Documents using a linear combination of BM25 and DPH. As our final model we propose Wikified + QE using entities to further improve the retrieval performance.
%\begin{math}
 %   \sum_{i\epsilon Q}\log\Bigg(\frac{N+0.5}{n{(i)}+0.5}\Bigg) \times \frac{(k_{1}+1).f_{i}}{K+f_{i}} \times\frac{(k_{2}+1).qf_{i}}{k_{2}+qf_{i}}
%\end{math}

\subsection{Data}

As per section \ref{sec:task}, we use the Podcast segment retrieval task to demonstrate usefulness of Wikipedia features in proactive information retrieval. The Spotify podcast Dataset contains approx 100,000 transcripts from around 60,000 hours of audio data culminating in the largest corpus of transcribed speech data. The episodes in the dataset were randomly sampled from 105,360 English podcast episodes published between January 1, 2019 to March 1, 2020 on Spotify with 10\% of the podcasts from professional creators with high production values and the other 90\% coming from amateurs \cite{clifton2020spotify}.
The data is also provided with training and testing topics (queries) along with human relevance judgements.  
There are 8 topics in the training dataset and another 50 topics in the test set.

To keep the computational costs low, we run the experiments outlined in section \ref{sec:experiments} exclusively using the training data in the podcast dataset. We first transform the dataset into two minute overlapping segments as described in \cite{jones2021trec}. Then, we take all the relevant segments as positive documents for the different queries. As there is a substantial number of segments that are irrelevant (as the whole dataset contains 3.5 Mn segments), we down sample this data and get a negative segment set of $\approx 14000$ segments that are not relevant to any of the labelled queries. This set of positive and negative segments make up the smaller scale dataset that we use in our empirical experiments. We refer to this dataset as \texttt{Podcast Small} dataset. We use the Wikifier \cite{wikifier} to associate Wikipedia concepts to both the queries (query + description) and the podcast segments as the models in section \ref{sec:models} require.  

\subsection{Experiments} \label{sec:experiments}

The empirical experiments we run in this work aim to answer RQ 1 and 2. We use the Podcast Small dataset for these experiments. In a proactive information retrieval setting, the user queries are inferred from users' historical interactions. We treat the query description provided in the dataset in the place of the features extracted from the user history. We hypothesise that the query description, can be used in two ways, i) as the user context that can be used in a zero effort query to make up the context and rank relevant documents from the corpus and/or ii) as a mean to disambiguate the true meaning of a short query provided by the user without prompting the user to provide further clarifications.     

To answer RQ1, we aim to see if there is a significant alignment between the Wikipedia concepts present in the query and a relevant document in contrast to a non-relevant document. To measure this quantitatively, we measure the Jaccard similarity coefficient $\rho$ between the query and the document using the sets of Wikipedia concepts. Then, for each query, we take the sets of values $\rho_{\text{rel}}$ and $\rho_{\text{non rel}}$ and compare the difference of medians using a one-tailed Mann–Whitney U test ($H_{alt.}:  \rho_{\text{rel}} > \rho_{\text{non rel}}$).

To answer RQ2, we take the annotated Wikipedia concepts of the query and the podcast segments as additional features. Then we develop a new set of information retrieval algorithms outlined in section \ref{sec:models} to account for the alignment of Wikipedia concepts present. Then we use the same Podcast Small dataset to evaluate if the ranking of relevant documents is different between the baselines and the proposed models.
To validate how Wikifying the context can be used to disambiguate a query, we ran Wikification on each query using i) the query words exclusively and ii) the query words + the description words, to observe if more precise identification of Wikipedia concepts is carried out. 

In order to run the evaluation experiments needed for RQ2, we utilise \texttt{Python Terrier} library \cite{Macdonald_2020}. The text processing and NER enrichment needed for the DCU model is done using the \texttt{Spacy} library. The Wikipedia annotations are sourced using the \texttt{Wikifier} service \cite{wikifier}. 

\subsubsection{Evaluation}
To evaluate if the Jaccard similarity coefficients in relation to RQ1 are significantly different, we use the test statistic of the hypothesis test. We present the performance of the ranking models related to validating RQ2 using NDCG, NDCG@30 and Precision@10 as these are the same evaluation metrics used in prior work \cite{jones2021trec}.
NDCG uses graded relevance instead of binary relevance/non-relevance to focus on highly relevant documents, implying that retrieving a highly relevant document is much more important than retrieving a less important document. Whereas precision measures the number of relevant documents retrieved over the total relevant documents. Similar to the prior work, we calculate NDCG for the entire set of relevant segments available and up-till rank 30. Precision is computed by using a cut-off of top-10 ranked segments.

\section{Experimental Results}

\begin{figure}[]
% \vskip 0.2in
\begin{center}
\centerline{\includegraphics[width=.37\columnwidth]{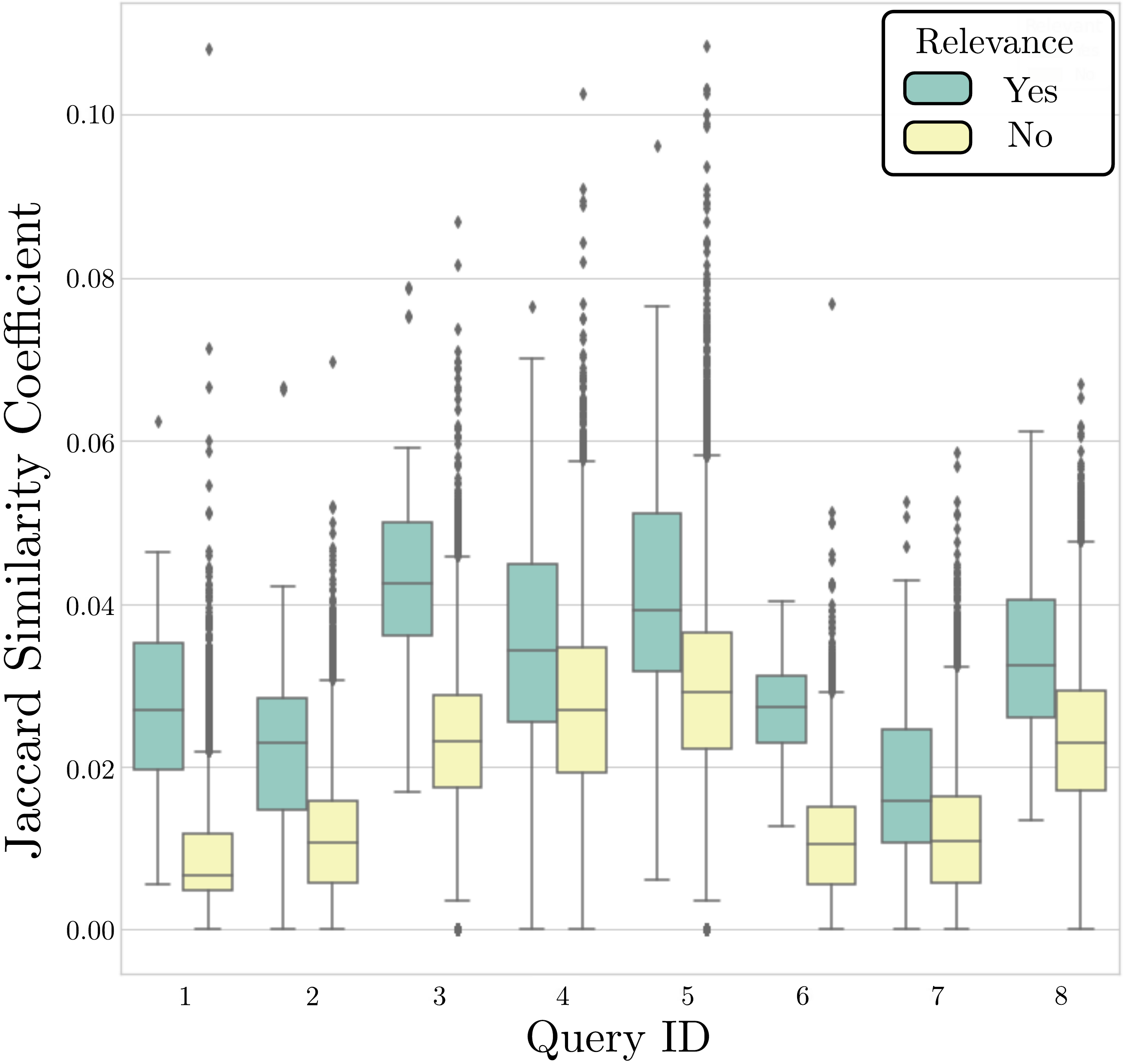}}
\caption{Jaccard similarity distribution of relevant vs. non-relevant segments for each query in the podcast dataset.} 
\label{fig:boxplots}
\end{center}
% \vskip -0.2in
\end{figure}

The results of the experiments are reported in this section. Figure \ref{fig:boxplots} shows the difference of medians of Jaccard similarity coefficients, while Table \ref{tab:jaccard} summarises the  results obtained from the hypothesis test conducted to verify RQ1. Table \ref{tab:ir} outlines the ranking performance obtained in the experiments run in order to verify RQ2. 

% \usepackage{graphicx}

\begin{table}[]
\caption{The difference of medians test for each query reported with the statistical significance.}
\label{tab:jaccard} \centering \small
\begin{tabular}{c|cc|ccc|cc}
\hline
\multicolumn{1}{l}{} & \multicolumn{2}{c}{Number of Documents} & \multicolumn{3}{c}{Median Jaccard   Similarity} & \multicolumn{2}{c}{MannWhitney U test} \\
Query ID             & Relevant         & Non-Relevant         & Relevant     & Non-Relevant     & Difference    & Test Statistic        & p value        \\
\hline
1                    & 70               & 14109                & 0.027        & 0.007            & 0.020         & 25680                 & 1.03E-49       \\
2                    & 63               & 14116                & 0.023        & 0.011            & 0.012         & 47466                 & 8.84E-46       \\
3                    & 78               & 14101                & 0.043        & 0.023            & 0.019         & 264688                & 1.22E-16       \\
4                    & 78               & 14101                & 0.034        & 0.027            & 0.007         & 393033                & 1.41E-06       \\
5                    & 80               & 14099                & 0.039        & 0.029            & 0.010         & 455015                & 1.42E-03       \\
6                    & 37               & 14142                & 0.027        & 0.010            & 0.017         & 35748                 & 8.37E-48       \\
7                    & 77               & 14102                & 0.016        & 0.011            & 0.005         & 56322                 & 2.80E-44       \\
8                    & 80               & 14099                & 0.032        & 0.023            & 0.010         & 271961                & 6.31E-16       \\
    \hline   
\end{tabular}
\end{table}

\begin{table}[]
\caption{Ranking performance of the models. The best performance in \textbf{bold} and the second best in \textit{italic} face.}
\label{tab:ir} \centering \small
    \begin{tabular}{l|ccc|ccc}
    \hline
    \multicolumn{1}{c}{}      &  \multicolumn{3}{c}{Features}  & \multicolumn{3}{c}{Metrics}   \\
    Model  & Query & NER        & Wiki         & NDCG & NDCG at 30 & Precision at 10 \\
           & Text  & Entities   & Concepts     &  & \\
    \hline
    \multicolumn{7}{c}{\emph{Baselines}}                  \\
    DPH            & $\surd$ & $\times$ & $\times$ & 0.48 & \textit{0.30}        & \textit{0.32}\\
    BM25           & $\surd$ & $\times$ & $\times$ & 0.48 & 0.28       & 0.31            \\
    DCU            & $\surd$ & $\surd$ & $\times$ & \textbf{0.51} & \textbf{0.32}       & 0.30             \\
    \hline
    \multicolumn{7}{c}{\emph{New Proposals}}              \\
    Wiki\_rel      & $\surd$ & $\times$ & $\surd$      & \textit{0.49} & 0.29       & {0.31}           \\
    Ent\_Wiki\_rel & $\surd$ & $\surd$ & $\surd$    & \textbf{0.51} & \textit{0.30}        & \textbf{0.36}           \\
    \hline
    \end{tabular}
\end{table}

% \begin{table}[]
% \caption{Information Retrieval on Baselines and QE Approaches using NER and Wikipedia }
% \label{tab:ir}
% \resizebox{\columnwidth}{!}{%
% \begin{tabular}{lllllllll}
% \cline{1-4}
% \multicolumn{1}{|l|}{\textbf{Model}} & \multicolumn{1}{l|}{\textbf{NDCG}} & \multicolumn{1}{c|}{\textbf{NDCG at 30}} & \multicolumn{1}{l|}{\textbf{Precision at   10}} & \textbf{} & \textbf{} & \textbf{} &  &  \\ \cline{1-4}
% \multicolumn{1}{|l|}{DPH} & \multicolumn{1}{l|}{0.48} & \multicolumn{1}{l|}{0.3} & \multicolumn{1}{l|}{0.32} &  &  &  &  &  \\ \cline{1-4}
% \multicolumn{1}{|l|}{BM25} & \multicolumn{1}{l|}{0.48} & \multicolumn{1}{l|}{0.28} & \multicolumn{1}{l|}{0.31} &  &  &  &  &  \\ \cline{1-4}
% \multicolumn{1}{|l|}{DCU NER QE} & \multicolumn{1}{l|}{0.51} & \multicolumn{1}{l|}{0.32} & \multicolumn{1}{l|}{0.3} &  &  &  &  &  \\ \cline{1-4}
% \multicolumn{1}{|l|}{Wikified} & \multicolumn{1}{l|}{0.49} & \multicolumn{1}{l|}{0.29} & \multicolumn{1}{l|}{0.31} &  &  &  &  &  \\ \cline{1-4}
% \multicolumn{1}{|l|}{Wikified+Entity} & \multicolumn{1}{l|}{0.51} & \multicolumn{1}{l|}{0.3} & \multicolumn{1}{l|}{0.36} &  &  &  &  &  \\ \cline{1-4}
% \end{tabular}%
% }
% \end{table}

\subsection{Query Keyword Disambiguation using the Context}

We hyptothesise that the context gathered by the user can also be used to disambiguate the meaning of keywords in a search query. To test this, we take 7 queries from the training dataset that belongs to categories \emph{topical} and \emph{known item} (''story about riding a bird'' is ignored as it is a query that belongs to the \emph{refinding} category). We enrich these queries using Wikification in two settings, i) query only vs. ii) query + description, to investigate if the identified Wiki concepts are different. Table \ref{tab:disam} reports how the salient entities in the query keywords are detected differently in the two settings. 

\begin{table}[]
\caption{The Quality of Query Word Disambiguation/Refinement when using the query exclusively vs. when using the description.}
\label{tab:disam} \centering \small
    \begin{tabular}{lll}
    \hline
    \multicolumn{1}{c}{Query}       & \multicolumn{1}{c}{Query Only}                       & \multicolumn{1}{c}{Query + Description}       \\
    \hline
    \underline{coronavirus}   spread            & wiki/Coronavirus      & wiki/Novel\_coronavirus \\
    \underline{greta   thunberg} cross atlantic & \multicolumn{1}{c}{-}                                & wiki/Greta\_Thunberg    \\
    \underline{black hole} image              & wiki/Black\_hole      & wiki/Black\_hole       \\
    \underline{daniel ek} interview           & \multicolumn{1}{c}{-}                                & wiki/Daniel\_Ek         \\
    \underline{michelle obama}  \underline{becoming}        & \multicolumn{1}{c}{-} & wiki/Michelle\_Obama    \\
                                    & wiki/Becoming\_(philosophy)                                & wiki/Becoming\_(book)  \\
    \underline{anna delvey}                   & wiki/Indian\_anna     & wiki/Anna\_Sorokin    \\
    \underline{facebook} \underline{stock}  prediction                   & wiki/Facebook         & wiki/Facebook          \\
                                    & wiki/Stock           & wiki/Stock             \\
    \hline
    \end{tabular}
\end{table}

\section{Discussion}

The results obtained in the preliminary experiments conducted are very promising. Majority of the results suggest that Wikipedia concepts can help the IR system.

\subsection{The Usefulness of Wikipedia Features (RQ1)}

Figure \ref{fig:boxplots} clearly shows that the Jaccard similarity coefficient between the relevant segments $\rho_\text{rel}$ and the non-relevant documents $\rho_\text{non rel}$ are different from each other. It is even evident that the confidence intervals don't even overlap in some of the queries in the training data. The test statistics presented in table \ref{tab:jaccard} further confirms this observation. In a one-tailed hypothesis test, all the queries demonstrate statistically significant differences of medians where the median Jaccard similarity of the Wikipedia concepts of the relevant segments and the query is greater than that of the non-relevant segments. These results enable us to conclude that the the overlap of Wikipedia concepts between queries and relevant documents is an influential signal that can be incorporated in relevance computation. The results indicate that there is promise in using Wikipedia concepts as a feature in IR systems where noise is present.

\subsection{Utility of Wikification in Information Retrieval (RQ2)}

The usefulness of Wikipedia concepts in relevance prediction is validated in the results presented in table \ref{tab:ir}. The results in the table show that replacing the NER-based entities with Wikipedia concepts is not as fruitful as augmenting them while the entities are retained. It is observed that both the proposed models outperform the entity based DCU model in precision at 10. This aligns with the our understanding of the effect of Wikipedia concepts. NER, while detecting specific persons, organisations and other entities in the text, only adds the textual representations of these entities to the query to expand it. Therefore, the newly added queries are not capable of entity disambiguation (when the same term occurs in a segment meaning something else. e.g. Apple, the fruit vs. the company) and coreference resolution (when a different textual form is used to mean the same entity. e.g. United Nations vs. UN). On the contrary, enriching both the query and the document with Wikipedia entities enable accounting for both of these effects, making the relevance calculation more precise. The \emph{Ent\_Wiki\_rel} model that contains both the NER-based entities and Wikipedia concepts, manages to improve the precision at 10 while retaining the NDCG performance. This shows that keeping the entities in the query is important to retain the NDCG gains found in the DCU model. 

In this scenario, we use the Wikipedia concepts and entities extracted from the description to improve the search ranking. While we treat the description in this scenario as a representation extracted from the user history, many concept-based user models described in section \ref{sec:lit} support extracting features such as keywords \cite{Syed2017} and Wikipedia concepts \cite{trueeducation} from content.   

\paragraph{Query Keyword Disambiguation} 

The results in table \ref{tab:disam} also give us insights into the potential of using query context for keyword disambiguation. It is seen that in multiple queries, certain concepts were not even detected when the query was used exclusively for Wikificaiton (e.g. Greta Thunberg, Michelle Obama etc.). However, with the presence of context (that can be extracted from historic encounter of concepts of a user \cite{truelearn}), the algorithm managed to identify these concepts in the query increasing recall. Table \ref{tab:disam} also shows instances where the precision of identification is improved in the presence of the context. The more precise \emph{Novel Coronavirus} was detected in the first query while the system understood that the query was about Michelle Obama's book, \emph{Becoming} in the fifth query, not the philosophical concept. These results show how many scenarios where short query keywords can be clarified automatically by referring to the context rather than needlessly prompting the human user.   

\subsection{Wikipedia Concepts and the User Experience}

\begin{figure} []
  \centering 
  \includegraphics[width=\linewidth]{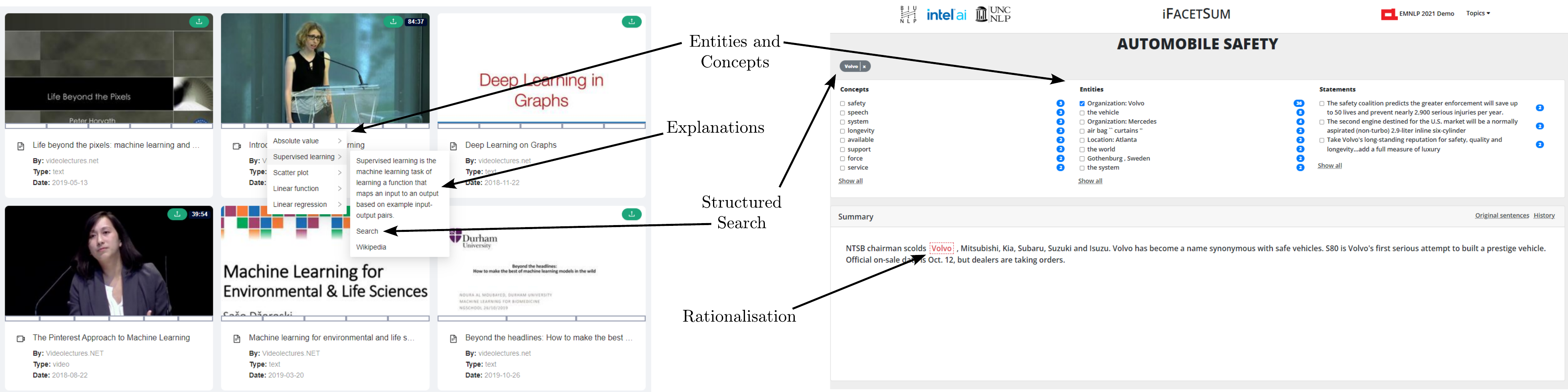}
  \caption{(i) X5Learn \cite{x5learn} and IFacetSum \cite{hirsch-etal-2021-ifacetsum} showing that using Wikipedia concepts to represent documents can support proactive information retrieval by summarising the key concepts/entities in the result collection, helping the user to make more informed decisions on their next steps.}
  \label{fig:ui}
\end{figure}

The previous section discussed one opportunity of Wikification, automatic query word disambiguation. Such features will substantially decrease the number of user prompts in a system leading to a smoother experience. Wikipedia concepts also bring the advantage of being grounded in a taxonomy that is intuitive to humans. The human intuitive nature of Wikipedia concepts allows the system to provide more transparent result sets that the user can navigate more efficiently. Figure \ref{fig:ui} shows two recent systems demonstrating the potential of using Wikipedia based representations to improve human-in-the-loop information retrieval.     

The left system, X5Learn \cite{x5learn} shows how different Wikipedia concepts are related to different segments of videos. The Jaccard-based relevance can be used to visualise the degree of concept overlap between the segments and the query indicating relevance (connecting to the hypothesis of RQ1). Also, using Wikipedia concepts rather than free-form keywords or NER-based entities allows the system to provide explanations of the concepts users encounter. This can be done using the knowledge base behind the taxonomy (in this case, Wikipedia as per figure \ref{fig:ui} (left)) Many users seeking information do not fully understand the knowledge landscape they navigate and can use learning opportunities to get more familiar with the topics.

IFacetSum, a multi-faceted mix-initiative interface shown in figure \ref{fig:ui}(right) is an example of how the result sets can be summarised using the entities and concepts in the document. As shown in the demonstration, the detected entities and concepts allow the user to understand the contents of the relevant documents quickly. Their user study also found that the users didn't utilise the statements and summaries as much as they used the concepts and entities \cite{hirsch-etal-2021-ifacetsum}. However, this system lacks the capabilities of explaining the topics as it is not backed by a knowledge base pointing to the advantages of using a taxonomy that has auxiliary information that can be used. In a nutshell, many recent visualisation approaches use concepts to represent documents. Having Wikipedia as a topic taxonomy enables these systems to make use of the advancements of Wikification that is advancing rapidly at present. It also allows taking use of other information in Wikipedia (e.g. semantic relatedness between topics, hierarchy of concepts etc.) to model the concepts in a much more sophisticated manner. This in turn, helps bridging the system seamlessly to the end user, creating cleaner interaction footprints that transform into the input for proactive search.     

\section{Conclusion}

Through this work, we explore the potential of using Wikipedia concepts to drive the proactive search experience of an IR system working on noisy documents. We explored the usefulness of Wikipedia concepts focusing on 2 research questions. We treated the query description in a podcast segment retrieval task dataset as the user context in a proactive IR system and ran a series of experiments. Our first experiment showed that the overlap of Wikipedia concepts between the relevant segments and the query are significantly larger than with the non-relevant segments. A subsequent experiment that added Wikipedia concept features to the search query and the document showed that the relevance model that uses Wikipedia concepts can improve precision of the ranking model while retaining the NDCG score in comparison to a counterpart that uses entities instead of Wikipedia concepts. A deeper analysis on the Wikification of the queries also showed that the meaning of the query keywords can be enhanced by using the query context. We also demonstrated using two recent works on how Wikipedia taxonomy can help an IR system to better connect with the end user and enable them to carry out structured IR tasks by proactively helping them navigate the information in the result sets. It is seen that having Wikipedia concepts allows presenting the result set in a humanly intuitive form that encourages the user to efficiently refine their query in the subsequent steps. All the observations from the analysed systems indicate that Wikipedia concepts have a key role to play in creating proactive information retrieval systems that can facilitate a higher degree human-in-the-loop operation. 
In the future work, we aim to use more sophisticated representations of Wikipedia concepts and advanced relevance models to run large scale experiments on the full podcast dataset with 3.5 Mn segments. Understanding how to use Wikipedia concepts effectively (e.g. by ranking them) is a top priority.   
% Many lines of future work exists from our findings. First, it is necessary to run the experiments outlined in section \ref{sec:experiments} with the full dataset that has 3.5 Mn podcast segments. We have also restricted our study to bag-of-word relevance models while models that use word embedding seem to show superior performance in this task \cite{jones2021trec}. There is potential to use the Wikipedia concept featues with dense word representations. We have barely scratched the surface in terms of \emph{how} we can incorporate Wiki features and compute relevance. More sophisticated Wikipedia representations and combining mechanisms should be tested in the future work. 

\begin{acknowledgments}
  This research was partially conducted as part of the X5GON project funded from the EU's Horizon 2020 research programme grant No 761758. This work is also supported by the European Commission-funded project ``Humane AI: Toward AI Systems That Augment and Empower Humans by Understanding Us, our Society and the World Around Us'' (grant 820437), EU Erasmus+ project ''European Network for Catalysing Open Resources in Education'' (project ref: 621586-EPP-1-2020-1-NO-EPPKA2-KA), and the AT2030 Programme. The AT2030 programme is funded by aid from the UK government and led by the Global Disability Innovation Hub. 
\end{acknowledgments}

%%
%% Define the bibliography file to be used
\bibliography{sample-ceur}

% \appendix


% The sources for the ceur-art style are available via
% \begin{itemize}
% \item \href{https://github.com/yamadharma/ceurart}{GitHub},
% % \item \href{https://www.overleaf.com/project/5e76702c4acae70001d3bc87}{Overleaf},
% \item
%   \href{https://www.overleaf.com/latex/templates/template-for-submissions-to-ceur-workshop-proceedings-ceur-ws-dot-org/pkfscdkgkhcq}{Overleaf
%     template}.
% \end{itemize}

\end{document}